\documentclass[10pt, conference]{IEEEtran}
\IEEEoverridecommandlockouts

\usepackage{CJKutf8}
\usepackage{cite}
\usepackage{amsmath,amssymb,amsfonts}
\usepackage{graphicx}
\usepackage{textcomp}
\usepackage{booktabs}
\usepackage{multirow}
\usepackage{here}
\usepackage{url}

\usepackage{graphicx}
\usepackage{textcomp}
\usepackage{xcolor}
\usepackage{orcidlink}
\usepackage{xparse}

\usepackage{scalerel}

\usepackage{tcolorbox}
\tcbuselibrary{breakable, skins, theorems}

\newtheorem{problem}{Problem}

\usepackage{cleveref}
\crefname{section}{Sec.}{Sec.}
\Crefname{section}{Sec.}{Sec.}
\crefname{figure}{fig.}{figs.}
\Crefname{figure}{Fig.}{Figs.}
\crefname{algorithm}{Alg.}{Alg.}
\Crefname{algorithm}{Alg.}{Alg.}
\crefname{problem}{Problem}{Problems}
\Crefname{problem}{Problem}{Problems}
\crefname{table}{table}{tables}
\Crefname{table}{Table}{Tables}

\usepackage{algorithm}
\usepackage[noend]{algpseudocode}

\usepackage{siunitx}

\newif\ifdraftComments
\def\mkDraftFn#1#2{%
  \expandafter\def\csname #1\endcsname##1{\ifdraftComments\textcolor{#2}{[#1: ##1]}\marginpar[$\longrightarrow$]{$\longleftarrow$}\fi}%
}
\mkDraftFn{RW}{red}

\NewDocumentCommand{\gone}{}{I}
\NewDocumentCommand{\gtwo}{}{I\hspace{-1.2pt}I}
\NewDocumentCommand{\gthree}{}{I\hspace{-1.2pt}I\hspace{-1.2pt}I}
\NewDocumentCommand{\gfour}{}{I\hspace{-1.2pt}V}
\NewDocumentCommand{\gfive}{}{V}
\NewDocumentCommand{\gsix}{}{V\hspace{-1.2pt}I}

\def\BibTeX{{\rm B\kern-.05em{\sc i\kern-.025em b}\kern-.08em
    T\kern-.1667em\lower.7ex\hbox{E}\kern-.125emX}}

\DeclareRobustCommand{\IEEEauthorrefmark}[1]{\smash{\textsuperscript{\footnotesize #1}}}

\makeatletter
\newcommand{\linebreakand}{%
  \end{@IEEEauthorhalign}
  \hfill\mbox{}\par
  \mbox{}\hfill\begin{@IEEEauthorhalign}
}

\makeatother

\begin{document}

\title{Online Job Scheduler for Fault-tolerant Quantum Multiprogramming\\
\thanks{
$\dagger$ These two authors contributed equally to this work.

This work was supported by
JSPS KAKENHI Grant Numbers JP22KJ1436, % 西尾DC1
25K21176, % 上野若手
New Energy and Industrial Technology Development Organization (NEDO), %西尾NEDO佐藤研
JST Support for Pioneering Research Initiated by the Next Generation: SPRING JPMJSP2110,
JST ACT-X JPMJAX23CT, % 脇坂さん
JST Moonshot R\&D JPMJMS226C, JPMJMS2061 , JPMJMS2067,
MEXT Q-LEAP JPMXS0120319794, JPMXS0118068682,
JST CREST JPMJCR23I4, JPMJCR24I4, 
JSPS Overseas Research Fellowships,
and the RIKEN Special Postdoctoral Researchers Program.}
}

\author{
    \IEEEauthorblockN{
    Shin Nishio\IEEEauthorrefmark{1,2}$^\dagger$\orcidlink{0000-0003-2659-5930},
    Ryo Wakizaka\IEEEauthorrefmark{3}$^\dagger$\orcidlink{0000-0001-8762-9335},
    Daisuke Sakuma\IEEEauthorrefmark{4}\orcidlink{0000-0002-1576-2185},
    Yosuke Ueno\IEEEauthorrefmark{5}\orcidlink{0000-0002-0402-9914},
    Yasunari Suzuki \IEEEauthorrefmark{5,6}\orcidlink{0000-0002-8005-357X},
    }
    \IEEEauthorblockA{\IEEEauthorrefmark{1}
    \textit{Graduate School of Science and Technology}, \textit{Keio University}, Yokohama, Kanagawa, Japan}
    \IEEEauthorblockA{\IEEEauthorrefmark{2}
    \textit{Department of Physics and Astronomy, University College London}, Gower Street, London, United Kingdom \\}
    \IEEEauthorblockA{\IEEEauthorrefmark{3}
    \textit{Graduate School of Informatics, Kyoto University}, Sakyo-ku, Kyoto, Japan}
    \IEEEauthorblockA{\IEEEauthorrefmark{4}
    \textit{Graduate School of Media and Governance, Keio University}, Fujisawa, Kanagawa, Japan}
    \IEEEauthorblockA{\IEEEauthorrefmark{5}
    \textit{Center for Quantum Computing, RIKEN}, Wako, Saitama, Japan}
    \IEEEauthorblockA{\IEEEauthorrefmark{6}
    \textit{Computer \& Data Science Laboratories, NTT Corporation}, Musashino, Tokyo, Japan}
}

\maketitle
\begin{abstract}
Fault-tolerant quantum computers are expected to be offered as cloud services due to their significant resource and infrastructure requirements. Quantum multiprogramming, which runs multiple quantum jobs in parallel, is a promising approach to maximize the utilization of such systems. 
A key challenge in this setting is the need for an online scheduler capable of handling jobs submitted dynamically while other programs are already running.

In this study, we formulate the online job scheduling problem for fault-tolerant quantum computing systems based on lattice surgery and propose an efficient scheduler to address it. To meet the responsiveness required in an online environment, our scheduler approximates lattice surgery programs, originally represented as polycubes, by using simpler cuboid representations. This approximation enables efficient scheduling while improving overall throughput. In addition, we incorporate a defragmentation mechanism into the scheduling process, demonstrating that it can further enhance QPU utilization.
\end{abstract}


\begin{figure*}[htbp]
    \centering
    \includegraphics[width=1\linewidth]{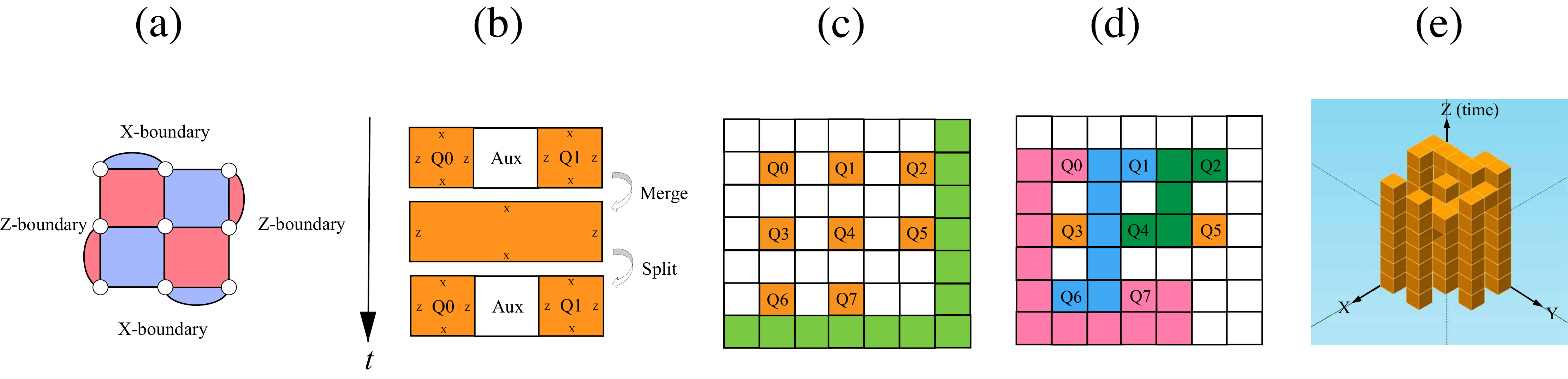}
    \caption{
    (a) A $[\![ d^2, 1, d ]\!]$ rotated surface code patch with $d=3$. The white circles represent the physical qubits for encoding the codeword. Red/blue regions are X/Z stabilizers of the code. 
    (b) Merging and splitting lattice surgery operations for realizing a two-qubit Pauli $ZZ$ measurement between $Q0$ and $Q1$ logical data qubits. Orange patches are logical data qubits and the white patch is a logical auxiliary qubit. $X/Z$ subscripts is showing the $X/Z$-boundary of the patch.
    (c) A floorplan of surface code patches. Green patches are frame qubits, which are used as auxiliary qubits for the lattice surgery paths.
    (d) Multiple lattice surgery operations were compiled on the floorplan of (c). Each color represents a gate. 
    (e) Representation of a quantum circuit with lattice surgery operations as a 3D polycube.}
    \label{fig:surface_code}
\end{figure*}

\section{Introduction}\label{sec:intro}

Quantum computers (QCs) have an advantage over classical computers in terms of the computational complexity for solving various problems~\cite{shor1994algorithms, grover1996fast}. 
In principle, however, the effect of noise cannot be ignored when executing large programs on QCs. 
To cope with this problem, fault-tolerant quantum computation (FTQC)~\cite{divincenzo1996fault, gottesman1998theory}, which periodically detects and corrects errors using quantum error-correcting codes (QECCs)~\cite{shor1995scheme, gottesman1997stabilizer}, has been proposed. 
Surface codes~\cite{bravyi1998quantum, dennis2002topological} are one of the QECCs and are expected to be utilized in FTQC because they are easy to implement on a two-dimensional (2D) qubit array and have a practical implementation of a two-qubit gate called lattice surgery~\cite{horsman2012surface}.

%%% (UENO) FTQCの計算様式について言及したい、physical -> logical levelのゲートとか、典型的な実行時間とか

%%% operating cost -> cost of ownership のほうがニュアンス近いですかね？
%The operating cost of quantum computers is high~\cite{parker2023estimating} since they require large-scale equipment and specialized control devices. 
QCs will likely be provided as cloud services since FTQC requires large-scale equipment and specialized control devices, and their cost of ownership is significantly high.
In particular, FTQC is expected to be especially costly because of the enormous overhead associated with the redundant encoded space and logical gates.
For example, Parker \textit{et al.}~\cite{parker2023estimating} estimated that the required electric power for superconducting QCs is approximately 125~MW, which costs \$$64,000$ for a single task of crypt analytics. Maximizing qubit utilization by running as many computational jobs in parallel as possible is necessary to obtain sufficient economic benefits, surpassing these high costs.
This approach is called \textit{quantum multiprogramming (QMP)}. 
Moreover, QMP becomes even more crucial for FTQC systems due to their substantial overhead; effectively utilizing logical qubits can significantly offset the high operational costs.
%%% TODO 特にFTQCのQMPが重要ということを言いたい
% This will result in high throughput.
% Such a system is called quantum multiprogramming.

% To obtain sufficient economic benefit bigger than these high running costs, it is required to increase the utilization of qubits by running as many calculation jobs in parallel as possible. This will result in a computation system with high throughput. Such a system is called quantum multiprogramming. There is strong motivation to use quantum multiprogramming in FTQC because the operational cost is expected to be high since it requires a huge overhead for the redundant encoded space and logical gates. 

% Although several job schedulers have been proposed for quantum multiprogramming in small-scale quantum computers without error correction, it is desirable to realize quantum multiprogramming in fault-tolerant quantum computers for the abovementioned reasons. 

% 上記のcostyな量子コンピュータは一部のクラウドベンダーや国研に集中して開発されることが初期は予測される。そしてユーザーはそれらの少数の集中型量子コンピュータを使うことになることが近い将来として予測される。マルチプログラミングは経済的利益だけではなく、ユーザーエクスピリエンスの向上とも両立し、実際に使い心地の良い(この表現がscientificかどうか微妙だな、、)量子コンピュータの実現には欠かすことができないと考えられる。https://dl.acm.org/doi/abs/10.1145/3352460.3358287?casa_token=0_XNxKWy_A4AAAAA:UCu_i5HucHYCsz4fZDyidetOo2h0OPhtAK8acc8bUY84xfDJtg8Trr2zSp9BhvNB4PyoM2ZfJlrffeY

% 量子誤り訂正を用いないNISQデバイスにおいては量子マルチプログラミング手法が提案されている。また、QOSはスケジューリングによってより高品質で待ち時間の少ないシステムを提供するが、これらのシステムはスケーラブルではないので、量子マルチプログラミングが真価を発揮するような大規模な量子計算システムには適さない。本仕事は、表面符号上の格子手術で記述される複数のジョブが適宜あたえられたとき、効率的に多くのジョブを処理する問題を定式化し、この問題を解くスケジューラを提案する。

To fully harness the benefits of QMP, appropriate job scheduling is essential.
Several QMP job scheduling techniques have been proposed, most of which focus on noisy intermediate-scale quantum (NISQ) devices without QECCs~\cite{das2019case, dou2020new, liu2021QuCloudNew, ohkura2022simultaneous, ravi2022AdaptiveJob, niu2023enabling, maurya2024understanding, giortamis2024qos}. 
NISQ-based QMPs focus on parallelizing physical-qubit-level operations. 
However, NISQ computers lack scalability and are therefore unsuitable for large-scale QC systems where the full potential of QMP can be realized.
Thus, it is imperative to develop job scheduling strategies specifically designed for the FTQC regime~\cite{weko_226758_1}.

It is important to note that FTQC differs from NISQ mainly in two key aspects: the representation and time scale of jobs. 
Because of these differences, existing NISQ-based QMP techniques cannot be applied directly to FTQC.

For the representation of FTQC jobs, lattice surgery is a well-studied one that facilitates operations on logical qubits encoded with surface codes by dynamically modifying their boundaries, effectively ``expanding'' and ``merging'' them.
Thus, in lattice surgery-based FTQC, each job is defined as a sequence of such expansions and merges of logical qubits on a 2D lattice.
This paradigm fundamentally differs from NISQ jobs, which focus on physical-qubit-level allocations and operations.

%%% execution time -> online scheduling 
The typical timescale for NISQ computation is within microseconds.
By contrast, in FTQC systems scaled to achieve quantum advantages, the execution time of jobs spans from several hours to days~\cite{yoshioka2024hunting}.
The scheduling of such long-duration jobs must prioritize responsiveness, which makes \textit{online job scheduling} techniques indispensable.
Online scheduling is essential in practical scenarios where jobs arrive unpredictably over time, requiring immediate decisions without full knowledge of future submissions.

In this work, we propose an online job scheduler for QMP with lattice surgery-based FTQC.
By incorporating appropriate preprocessing, our job scheduler can effectively manage dynamically submitted FTQC jobs with high responsiveness.
In addition, by adapting the concept of system fragmentation, which was originally studied in HPC job scheduling~\cite{feitelson1997job, mu2002utilization}, to the FTQC context, we enhance the logical qubit utilization.
Our online scheduler employs dynamic reallocation of jobs to reduce system fragmentation (\textit{defragmentation}), thereby improving overall system utilization and total job throughput.

Our contributions are summarized as follows:
\begin{enumerate}
\item We formalize the online job scheduling problem for QMP in FTQC. (Sec.~\ref{sec:online-schedule})
\item We propose preprocessing before scheduling to improve the responsiveness of the job scheduler. (Sec.~\ref{sec:preprocess})
\item We propose greedy- and integer linear programming (ILP)-based online job schedulers. (Sec.~\ref{sec:ILP-scheduler},\ref{sec:greedy-scheduler})
\item We introduce defragmentation to our schedulers for further improvements in the throughput. (Sec.~\ref{sec:defrag})
\item Our evaluation demonstrates that the proposed scheduler, together with the defragmentation technique, improves the throughput by $2.3$ to $2.4$ times on average, with a maximum improvement of $4.53$ times. (Sec.~\ref{sec:evaluation})
\end{enumerate}

% \begin{figure*}[htbp]
%     \centering
%     \includegraphics[width=1\linewidth]{figs/surfacecode.pdf}
%     \caption{
%     (a) A $[\![ L^2, 1, L ]\!]$ rotated surface code patch with $L=3$. The white circles represent the physical qubits for encoding the codeword. Red/blue regions are X/Z stabilizers of the code. 
%     (b) Merging and splitting lattice surgery operations for realizing $CX$ gate between $Q0$ and $Q1$ logical data qubits. Orange patches are logical data qubits and the white patch is a logical auxiliary qubit. $X/Z$ subscripts is showing the $X/Z$-boundary of the patch.
%     (c) A floorplan of surface code patches. Green patches are frame qubits, which are used as auxiliary qubits for the lattice surgery paths.
%     (d) Compiled multiple lattice surgery operations on the floorplan of (c). Each color represents a gate. 
%     (e) Representation of a quantum circuit with lattice surgery operations as a 3-dimensional polycube.}
%     \label{fig:surface_code}
% \end{figure*}
\section{Preliminary}\label{sec:background}

In this section, we give a brief overview of FTQC with lattice surgery on surface codes.
QCs are fundamentally unable to ignore the effects of errors. 
To address this, FTQC\cite{divincenzo1996fault, gottesman1998theory} has been proposed, in which QECCs are used to periodically correct errors while performing computations. This period is called a code cycle and used as a time unit of FTQCs.

Surface codes are considered one of the most promising candidates for FTQC. 
They can be implemented on a 2D planar without remote interactions. 
Moreover, they support multi-qubit logical gate implementations within the 2D constraints, using techniques such as lattice surgery\cite{horsman2012surface} and defect braiding\cite{fowler2012surface}. 
In addition, surface codes have a high noise threshold, enabling a lower error rate for logical qubits compared to physical qubits without encoding in a realistic error rate region.
% Planar surface codes are quantum error-correcting codes defined on a 2D lattice, requiring only two-qubit interactions between neighboring qubits.
\emph{(Rotated) surface codes} are QECCs defined on a $45^\circ$-rotated 2D lattice, requiring only two-qubit interactions between neighboring qubits. 
A rotated surface code encodes one logical qubit into $d^2$ physical qubits with code distance $d$, where $d$ is the lattice size. These parameters are denoted by $[\![ d^2, 1, d ]\!]$. Fig.~\ref{fig:surface_code}(a) shows a rotated surface code with $d=3$.

Lattice surgery is a technique used to implement logical gates on surface codes. 
It enables operations such as entangling gates and measurements by merging and splitting patches of the surface code. 
These operations are performed by temporarily modifying stabilizer measurements along the shared boundaries of code patches.
For example, a logical CNOT gate between two encoded qubits can be realized through a sequence of lattice surgery operations involving joint stabilizer measurements across the boundary of two patches. 
Fig.~\ref{fig:surface_code}(b) shows a two-qubit Pauli $ZZ$ measurement implemented via merging and splitting operations. 
This method supports scalable and efficient quantum computation while preserving fault tolerance. 
Then, a CNOT gate can be realized by a zigzag path connecting control and target qubit patches across auxiliary qubit patches.

Lattice surgery is particularly appealing because it reduces space overhead compared to other approaches and is well-suited for 2D architectures with nearest-neighbor interactions.
To implement a given quantum circuit using lattice surgery operations, surface code patches need to be arranged on a 2D plane with auxiliary surface code patches.
This arrangement is referred to as a \emph{floorplan}.
Various floorplans with different fill rates have been proposed~\cite{beverland2022surface, chamberland2022universal, ueno2024high, lee2021even, kobori2024lsqca}.
Fig.~\ref{fig:surface_code}(c) illustrates a floorplan with $8$ data-qubit patches with surface codes. 

For example, a $25$\% fill rate floorplan on $W \times H$ 2D grid~\cite{beverland2022surface} has been studied well. We later utilize this floorplan for our benchmark, while our QMP methods can be used for any floorplan.
In addition to the $25$\% fill rate region, we add one row and one column of auxiliary qubits to the floorplan, as shown as the green region in Fig.~\ref{fig:surface_code}(c). 
We refer to these added qubits as \emph{frame qubits}. 
Frame qubits are particularly useful in high-fill-rate floorplans, as they guarantee that lattice surgery operations can be applied between any qubits facing the frame.
% In this work, we use a floorplan where the patches are placed on a $W \times H$ two-dimensional grid with a $25$\% fill rate\cite{beverland2022surface}. 

Given a floorplan, a quantum circuit can be compiled into a series of zigzag lattice surgery operations mapped onto the floorplan.
Fig.\ref{fig:surface_code}(d) shows compiled gates on qubit pairs (Q0, Q7), (Q1, Q6), and (Q2, Q4) using the floorplan in Fig.\ref{fig:surface_code}(c).

Finally, the entire quantum circuit can be compiled and represented as a sequence of lattice surgery operations. 
If one code cycle is considered a unit of time, the sequence can be visualized as a \emph{polycube}, where each cube has height and width equal to the code distance $d$, and the depth corresponds to time steps.
Fig.\ref{fig:surface_code}(e) illustrates a polycube for a quantum circuit with $8$ data qubits on the floorplan from Fig.\ref{fig:surface_code}(c).
\section{Related Work}\label{sec:related}

\subsection{Multiprogramming in NISQ}

In the context of NISQ, various multiprogramming techniques have been proposed to date~\cite{das2019case,
dou2020new,
liu2021QuCloudNew,
ohkura2022simultaneous,
ravi2022AdaptiveJob,
niu2023enabling,
maurya2024understanding,
giortamis2024qos}. A central challenge in NISQ-based QMP is improving throughput.

Recent research on QMP has branched into several distinct but complementary directions: 

(1) development of scheduling algorithms that manage when and how jobs are executed on shared quantum hardware~\cite{das2019case, liu2021QuCloudNew, ohkura2022simultaneous, ravi2022AdaptiveJob,niu2023enabling, giortamis2024qos}. These works aim to optimize latency, maximize qubit utilization, and reduce idle time across workloads.

(2) addressing the challenge of allocating physical qubits to multiple programs while minimizing interference and communication overhead~\cite{dou2020new, liu2021QuCloudNew, ohkura2022simultaneous, niu2023enabling, giortamis2024qos}. These approaches develop mapping strategies that ensure spatial separation and reduce crosstalk among co-located circuits.

(3) system-level solutions for supporting QMP in cloud environments~\cite{liu2021QuCloudNew, ravi2022AdaptiveJob, niu2023enabling, giortamis2024qos}. These include job isolation mechanisms, runtime scheduling services, and even proposals for full-stack quantum operating systems that manage concurrency and user separation~\cite{giortamis2024qos}.
% while mitigating fidelity degradation caused by noise~\cite{dou2020new} and crosstalk errors~\cite{ohkura2022simultaneous}. 

% To address this, several techniques have been introduced, including (1) fair allocation of high-fidelity qubits across jobs~\cite{das2019case}, (2) insertion of inter-program SWAP gates to reduce communication and SWAP overhead~\cite{liu2021QuCloudNew}, and (3) adaptive strategies that monitor fidelity and dynamically dispatch jobs~\cite{das2019case,ravi2022AdaptiveJob}.

In addition to those works, a security vulnerability in QMP caused by information leakage from a job to neighboring jobs~\cite{maurya2024understanding, giortamis2024qos} has been pointed out and studied.

QMP techniques for FTQC differ fundamentally from those designed for NISQ. Most notably, the scheduler is generally not required to account for noise, except in the case of hardware faults. Moreover, the extended cycle time introduced by error-correction decoding allows for the implementation of more sophisticated job scheduling algorithms than is feasible in NISQ systems. In addition, new abstractions for managing shared resources, such as magic states and memory-dedicated regions~\cite{kobori2024lsqca}, will be essential. This work takes a first step toward realizing FTQC-based QMP.

\subsection{Multiprogramming in FTQC}
Devitt~\textit{et al.} have introduced an FTQC architecture shared by multiple users\cite{devitt2008high}. They propose to generate a large 3D cluster state called a global lattice, which will be divided for each job, enabling QMP.  

To our knowledge, Nakayama \textit{et al.}~\cite{weko_226758_1} first introduced a job scheduler for FTQC-based QMP. Their study, like ours, focuses primarily on lattice surgery in surface codes and proposes a scheduler based on integer linear programming. However, their proposed method differs from ours in that it is an offline scheduler based on an exact polycube representation. In addition, they did not assess the responsiveness of the scheduler, leaving a significant gap in making it operational for real-world systems. As we will discuss in Sec.\ref{sec:evaluation}, applying their scheduler in an online setting fails to improve the utilization of quantum processors. In contrast, our schedulers achieve both higher responsiveness and better utilization of quantum processors. Additionally, unlike our approach, their study does not consider reallocating or transforming already assigned jobs to improve further the utilization of quantum processors, which is a key aspect of our method.
% Memo
% - 直方体表現は使う領域と実行時間を事前に指定するような状況に相当
\section{Online Scheduler For Fault-Tolerant Multiprogramming}\label{sec:online-schedule}

As described in \Cref{sec:background}, a lattice surgery program can be represented as a polycube. Based on this representation, we define the online job scheduling problem for lattice surgery programs as follows.

\begin{tcolorbox}
\begin{problem}[Online scheduling]\label{prob:online-scheduling}
Given a program sequence $P_1, \dots, P_n$, where $P_i$ is the $i$-th request represented by a polycube. The goal is to find the schedule $S_1, \dots, S_n$ that minimizes the total execution time $\max_i \{ t_{\rm fin}(S_i(P_i)) \}$ where each schedule $S_i$ contains the schedule information, which consists of the $3\rm D$ positions, rotation and whether the program will be flipped. We denote by \( t_{\mathrm{fin}}(P) \) the time at which program \( P \) completes its execution.
\end{problem}
\end{tcolorbox}

\begin{figure}[htbp]
    \centering
    \includegraphics[width=0.6\linewidth]{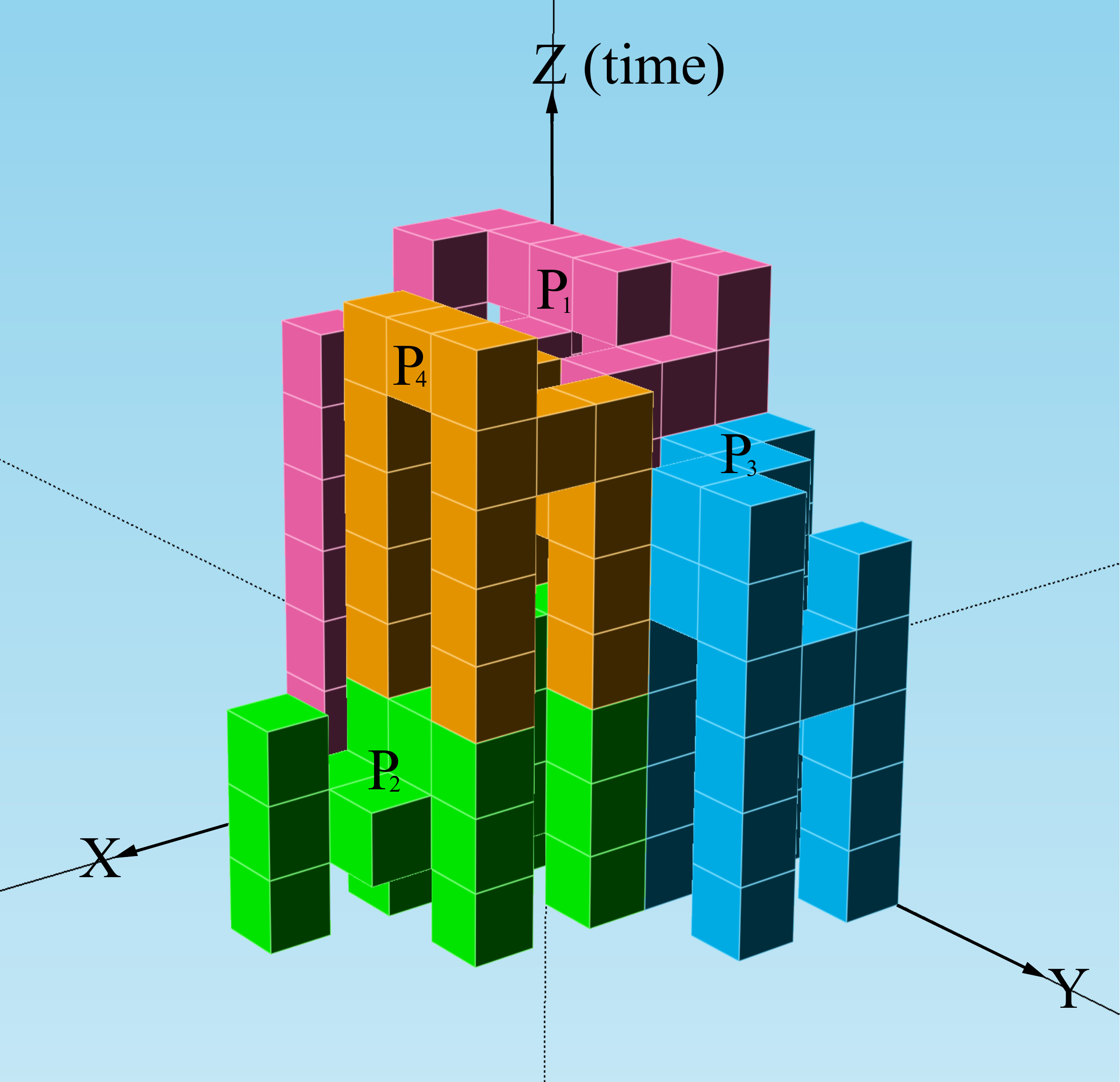}
    \caption{An example of schedule for $\mathcal{P}_1, \dots, \mathcal{P}_4$}
    \label{fig:instance}
\end{figure}

\begin{figure*}[tb]
    \centering
    \includegraphics[width = 0.7\linewidth]{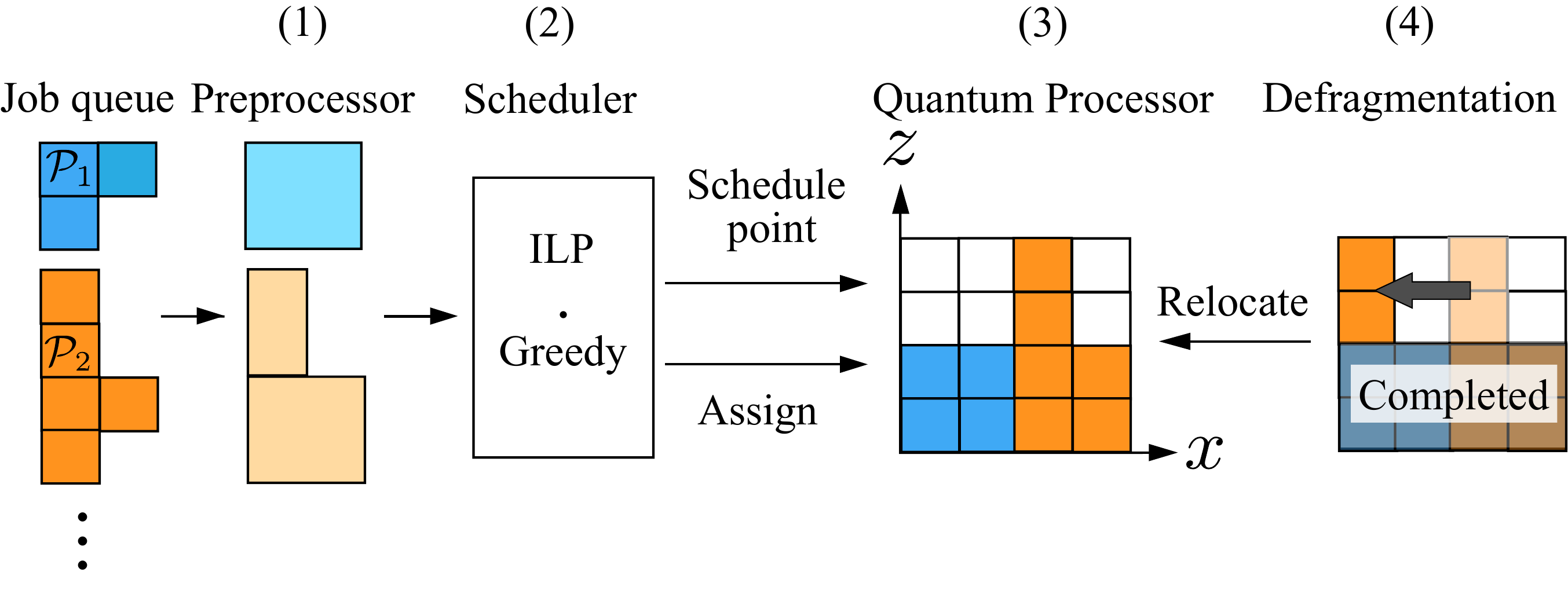}
    \caption{The workflow of the scheduling for FTQC Multiprogramming}
    \label{fig:workflow}
\end{figure*}

We show an example of a schedule for a randomly generated instance of Problem $1$ in Fig.~\ref{fig:instance}.

\Cref{prob:online-scheduling} is clearly more complex than the 3D bin packing problem (3D-BPP) and is therefore NP-hard. Accordingly, this study proposes a scheduler based on heuristic methods.

\subsection{Overview of Our Scheduling Workflow}\label{sec:overview}
We address \cref{prob:online-scheduling} using the workflow illustrated in \Cref{fig:workflow}. Our scheduling workflow consists of (1) preprocessor, (2) scheduler, (3) quantum processor, and (4) defragmentation.

\subsubsection{Preprocessor (\ref{sec:preprocess})}
The preprocessor can transform the program representation. In this study, we specifically use this approach to simplify the complex polycube representation, thereby accelerating the scheduling process. 

\subsubsection{Scheduler (\ref{sec:ILP-scheduler}, \ref{sec:greedy-scheduler})}
The scheduler takes several jobs from the job queue to be scheduled to the quantum processor. This study proposes two types of schedulers: an ILP-based scheduler and a corner-greedy scheduler.

Before starting each job scheduling process, the scheduler notifies the system of a \emph{schedule point} \( t \). The schedule point represents the earliest possible time at which the scheduler may assign a job. If the scheduler fails to return a scheduling result by the schedule point, the system pauses the entire program execution at \( t \) and waits for the scheduler's response. This mechanism ensures that scheduling and the execution of already allocated programs can safely proceed in parallel. Such a protocol is particularly important when using scheduling methods like ILP-based schedulers, where the scheduling time is difficult to predict precisely.  

In this study, we set the schedule point by notifying the system of a future time offset from the current execution point. This offset is determined by the average number of code cycles required for past scheduling operations.

Additionally, this study adopts a batch-processing approach. Specifically, for each scheduling cycle, up to \( B \) jobs (where \( B \) is the batch size) are extracted from the job queue and scheduled simultaneously. We have to choose the batch size appropriately based on the scheduler's desired responsiveness.

\subsubsection{Quantum Processor}

The quantum processor is represented as an infinitely tall cuboid \([0, W) \times [0, H) \times [0, \infty)\). Here, \( W \) and \( H \) correspond to the chip's width and height, respectively, while the \( z \)-axis represents the time dimension.

%The following \cref{sec:preprocess,sec:ILP-scheduler,sec:greedy-scheduler} provide a detailed explanation of the preprocessing and scheduler proposed in this study.

\subsubsection{Defragmentation}
Defragmentation (\Cref{sec:defrag}) is the process of relocating scheduled jobs to create larger contiguous blocks of free logical qubit patches, thereby enabling the placement of new job requests. \cref{sec:defrag} provides a detailed description of the defragmentation algorithm.

\subsection{Preprocessing: Conversion to Cuboid Representation}\label{sec:preprocess}

When processing user-submitted jobs in an online manner, it is necessary to balance minimizing scheduling time while searching for the most optimal placement possible. However, search methods based on the polycube representation, which involves complex geometric shapes, are computationally expensive. This increased cost reduces the scheduler's responsiveness, potentially limiting the extent of the search and leading to suboptimal scheduling decisions.

\begin{figure}[htbp]
  \centering
  \includegraphics[width=0.7\linewidth]{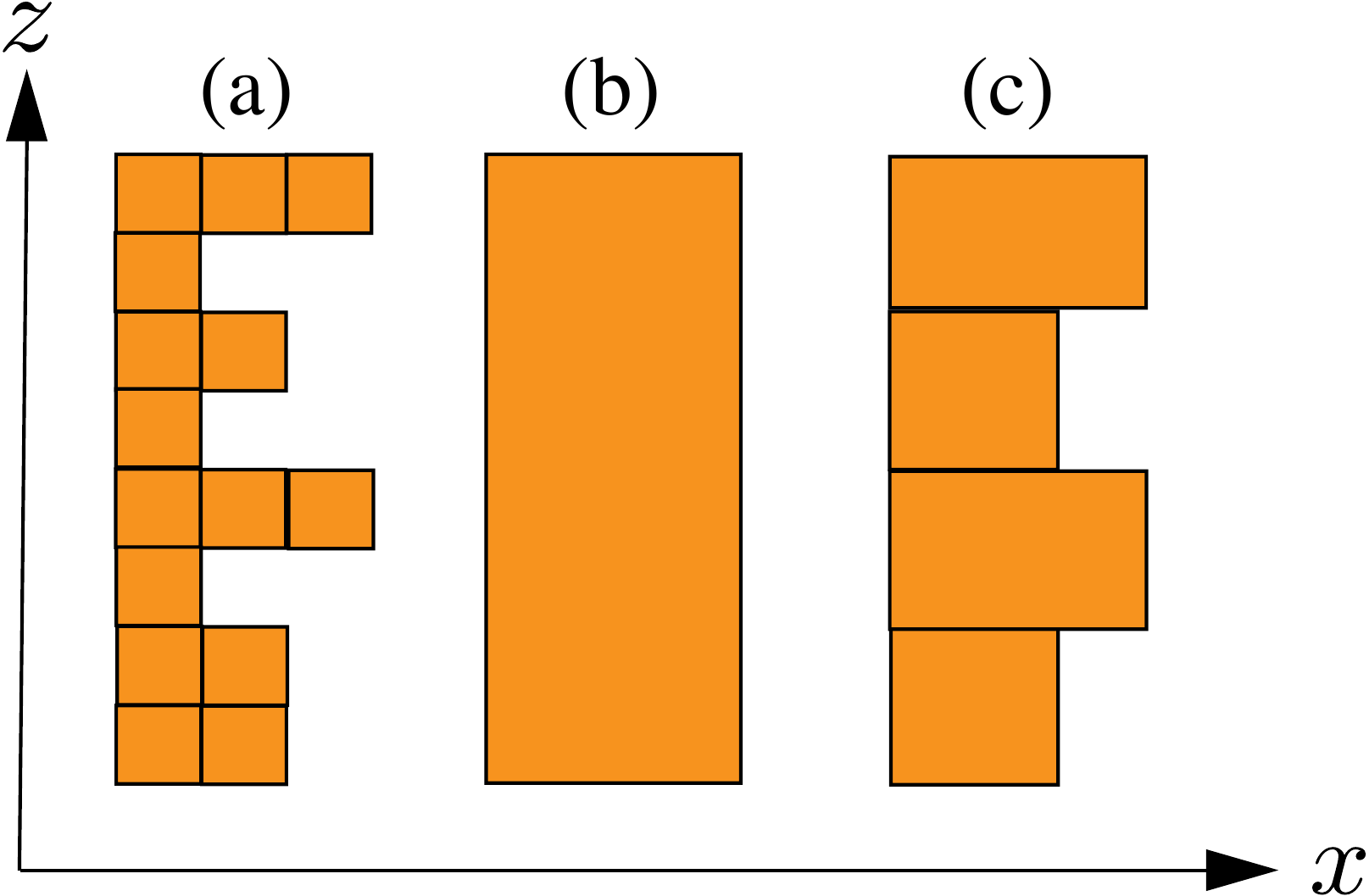}
  \caption{Preprocessing of a polycube. For simplicity, the illustration is shown in two dimensions. (a) A polycube. (b) A cuboid which is a bounding box of (a). (c) $k$-cuboids. Original polycube divided into $k$ segments perpendicular to the time axis, each approximated by a bounding box.}
  \label{fig:k-split}
\end{figure}

To address this issue, we introduce a preprocessing step that simplifies the polycube representation into a cuboid representation as shown in \Cref{fig:k-split}(a) and (b). This cuboid representation is defined as the minimum bounding box of the polycube. This transformation reduces \Cref{prob:online-scheduling} to a simpler online 3D bin packing problem (3D-BPP)~\cite{martello2000ThreeDimensionalBin,zhao2022Online3D}, enabling more responsive scheduling.

If the chip area occupied by a program is changed during execution, a more generalized \( k \)-cuboid representation can be considered as shown in \Cref{fig:k-split}(c). This is obtained by dividing the original polycube along the \( z \)-axis into \( k \) segments and computing the minimum bounding box for each segment. Although the scheduler proposed in this study can be extended to support \( k \)-cuboid representations, we assume in the following discussion that all programs are represented using the \( 1 \)-cuboid form.

\subsection{ILP-Based Scheduler}\label{sec:ILP-scheduler}

Let \( P_1, \dots, P_n \) be the programs (cuboids) to be scheduled. To formulate the ILP problem, we introduce the following constants and variables.

\begin{itemize}
\item \( W, H \) represents the size of the quantum processor.  
\item \( w_i, h_i, l_i \) denote the width, height and length of \( P_i \), respectively.
\item The total execution time constraint for all programs is set as \( L := \sum_i l_i \).  
\item $a_{ij}, b_{ij}, c_{ij}$ are decision variables, which becomes 1 if and only if the program $P_i$ is positioned before $P_j$ and the cuboid projected to $x$, $y$, and $z$-axis do not overlap, respectively.
\item \( x_i, y_i, z_i \) are non-negative integer variables representing the coordinates of program \( P_i \).  
\item \( v \) is the objective variable representing the number of cycles required for all programs to complete execution.
\end{itemize}

Using these definitions, the scheduling problem can be formulated as the following ILP problem~\cite{chen1995AnalyticalModel}:

\begin{tcolorbox}[left=0.4em]
\begin{problem}[ILP problem based on cuboids]\label{prob:ILP}
\begin{align}
  &\textrm{\rm minimize} \quad v    \\
  &\textrm{\rm subject to} \\
  &\quad \forall i, j \in \{1, \dots, n\}. \notag\\
  &\qquad a_{ij} + a_{ji} + b_{ij} + b_{ji} + c_{ij} + c_{ji} \geq 1 \\
  &\qquad w_i + W(a_{ij} - 1) \leq x_j - x_i \label{eq:x-rel-pos} \\
  &\qquad h_i + H(b_{ij} - 1) \leq y_j - y_i \\
  &\qquad l_i + L(c_{ij} - 1) \leq z_j - z_i \\
  &\quad \forall i \in \{1 \dots, n\}. \notag\\
  &\qquad x_i + w_i \leq W \label{eq:x-size} \\
  &\qquad y_i + h_i \leq H \label{eq:y-size} \\
  &\qquad z_i + l_i \leq L \label{eq:z-size} \\
  &\qquad z_i + l_i \leq v
\end{align}
\end{problem}
\end{tcolorbox}

Equation~(\ref{eq:x-rel-pos}) can be rewritten based on the values of \( a_{ij} \) as follows:
\begin{align}
  &w_i \leq x_j - x_i, & (a_{ij} = 1) \label{eq:a1}\\
  &x_i - x_j + w_i \leq W. & (a_{ij} = 0) \label{eq:a0}
\end{align}  
Equation~(\ref{eq:a1}) ensures that the \( j \)-th cuboid is positioned at least \( X_i \) units away from the \( i \)-th cuboid, effectively preventing overlap in the \( x \)-axis. On the other hand, (\ref{eq:a0}) follows automatically from (\ref{eq:x-size}) and does not impose any additional meaningful restrictions, meaning it does not influence the solution.  

Thus, the decision variable \( a_{ij} \) successfully determines whether two cuboids in the schedule overlap in the \( x \)-axis. The same reasoning applies to the \( y \) and \( z \) coordinates, ensuring proper collision detection across all axes.

Actually, the placement of previously assigned jobs must also be taken into account when scheduling new jobs. This can be modeled by adding cuboids in which the variables \( x_i \), \( y_i \), and \( z_i \) are fixed constants, and the constraints (\ref{eq:x-size}), (\ref{eq:y-size}), and (\ref{eq:z-size}) can be ignored since they are already satisfied.

The number of decision variables \( a_{ij}, b_{ij}, c_{ij} \) is \( \mathcal{O}(n^2) \) and does not depend on the size of the programs (i.e., $W$, $H$, $L$). In contrast, the method based on the polycube representation proposed by Nakayama \textit{et al.}~\cite{weko_226758_1} requires decision variables to represent the exact placement of programs in the space, making their number $\mathcal{O}(W \times H \times L)$. When handling large-scale programs, a formulation like ours, which is independent of the spatial resolution, is preferable for efficiency and scalability.

\subsection{Corner Greedy Scheduler}\label{sec:greedy-scheduler}

Although the ILP-based scheduler proposed in the previous section has the potential to explore optimal placements, it faces challenges in terms of responsiveness. In scenarios where scheduling a larger number of jobs efficiently is a priority, a near-optimal scheduler that sacrifices optimality for improved responsiveness is required.

To implement a highly responsive scheduler, it is essential to determine the placement of each incoming job quickly. In this study, we propose a scheduler that searches for placement candidates by targeting the corners of previously allocated jobs~\cite{martello2000ThreeDimensionalBin}. The scheduler algorithm is given in \cref{alg:corner-greedy}. In the algorithm, the coordinates of a cuboid \( P \) are denoted as \( P.x_i, P.y_i, P.z_i \) for \( i = 1, 2 \), where \( P.x_2 = P.x_1 + P.\mathrm{size}_x \).

\begin{figure}[tb]
    \centering
    \includegraphics[width=0.5\linewidth]{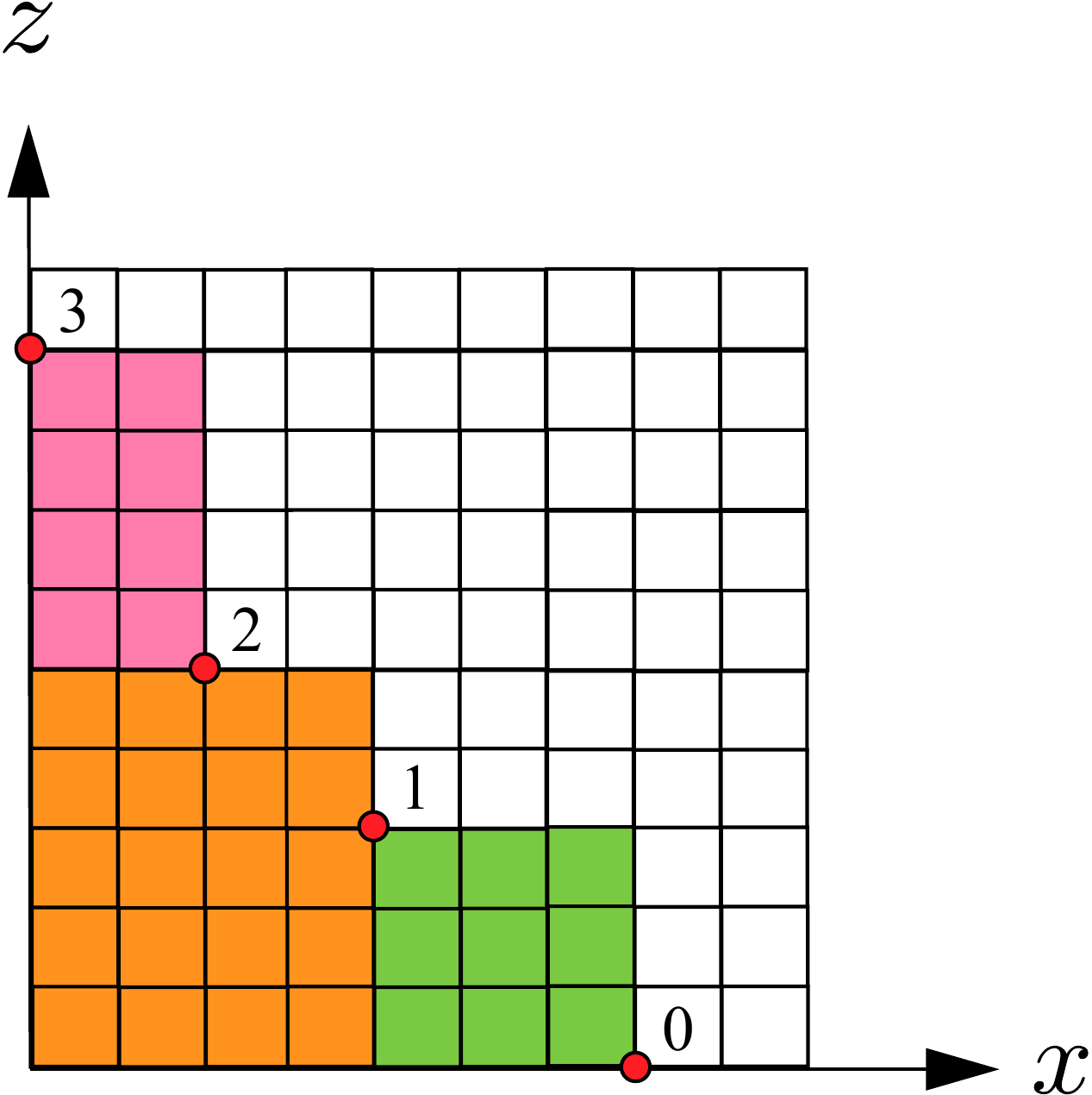}
    \caption{An example of potential placement candidates for the corner-greedy scheduler. Each colored box is a scheduled job, and red dots are the candidates.}
    \label{fig:corner-greedy-2D}
\end{figure}

Our strategy is based on the intuition that, when placing a new cuboid, it is efficient to position it near the corners of previously placed cuboids. More precisely, if a cuboid was placed at \((x_1, y_1, z_1)\), then potential placement candidates for the next cuboid would include the points $(x_2, y_1, z_1), (x_1, y_2, z_1)$ and $(x_1, y_1, z_2)$. Fig.~\ref{fig:corner-greedy-2D} illustrates such candidate corner points in the case where the problem is restricted to two dimensions. Additionally, to ensure that a feasible solution always exists, the coordinate \( (0, 0, z_2) \) is included as a candidate point. This guarantees the presence of a candidate point \( (0, 0, z_2) \) at which any cuboid can be placed.

Our scheduler selects, among the candidate positions, the one with the smallest \( z \)-coordinate that does not conflict with other programs. If multiple valid positions with the minimal \( z \) exist, the scheduler chooses the position that minimizes \( x + y \), thereby placing the program as close as possible to the origin of the \( xy \)-plane. This method allows for efficient space utilization while maintaining a computationally lightweight scheduling process.

\begin{figure}[!t]
  \begin{algorithm}[H]
    \caption{Corner-greedy algorithm}
    \label{alg:corner-greedy}
    \begin{algorithmic}[1]
      \Require{$L$ is the set of location candidates to be assigned}
      \Require{$\mathcal{P_{\text{rsv}}}$ is the set of the reserved jobs}
      \Function{CornerGreedySchedule}{$L$}
      \State ${\rm sp} \gets \Call{EstimateSchedulePoint}{ }$
      \State $\mathcal{P} \gets \Call{TakeWaitingJobs}{B}$
      \Comment{$B$ is the batch size}
      \State $L \gets \{(x, y, z) \in L \mid \mathrm{sp} \leq z \}$
      \For{$P \in \mathcal{P}$}
        \State $P_{\text{best}} \gets \bot$
        \For{$(x, y, z) \in L$}
          \State $P' \gets \Call{Assign}{P, (x, y, z)}$
          \If{$P'$ does not overlap with $\forall Q \in \mathcal{P}_\text{rsv}$}
            \If{$P_{\text{best}} = \bot \lor P'.z_1 < P_{\text{best}}.z_1 \lor (P'.z_1 = P_{\text{best}}.z_1 \land P'.x_1 + P'.y_1 < P_{\text{best}}.x_1 + P_{\text{best}}.y_1)$}
              \State $P_{\text{best}} \gets P'$
            \EndIf
          \EndIf
        \EndFor
        \State $(x_1, y_1, z_1) \gets (P_{\text{best}}.x_1, P_{\text{best}}.y_1, P_{\text{best}}.z_1)$
        \State $(x_2, y_2, z_2) \gets (P_{\text{best}}.x_2, P_{\text{best}}.y_2, P_{\text{best}}.z_2)$
        \State $L \gets L \setminus \{ (x_1, y_1, z_1) \}$
        \State $L \gets L \cup \{ (x_1, y_1, z_2) , \ldots, (x_2, y_1. z_1), (0, 0, z_2) \}$
        \State $\mathcal{P}_{\text{rsv}} \gets \mathcal{P}_{\text{rsv}} \cup P_{\text{best}}$
      \EndFor
      \EndFunction
    \end{algorithmic}
  \end{algorithm}
\end{figure}

\section{Defragmentation}\label{sec:defrag}

\begin{figure*}[tb]
    \centering
    \includegraphics[width=0.8\linewidth]{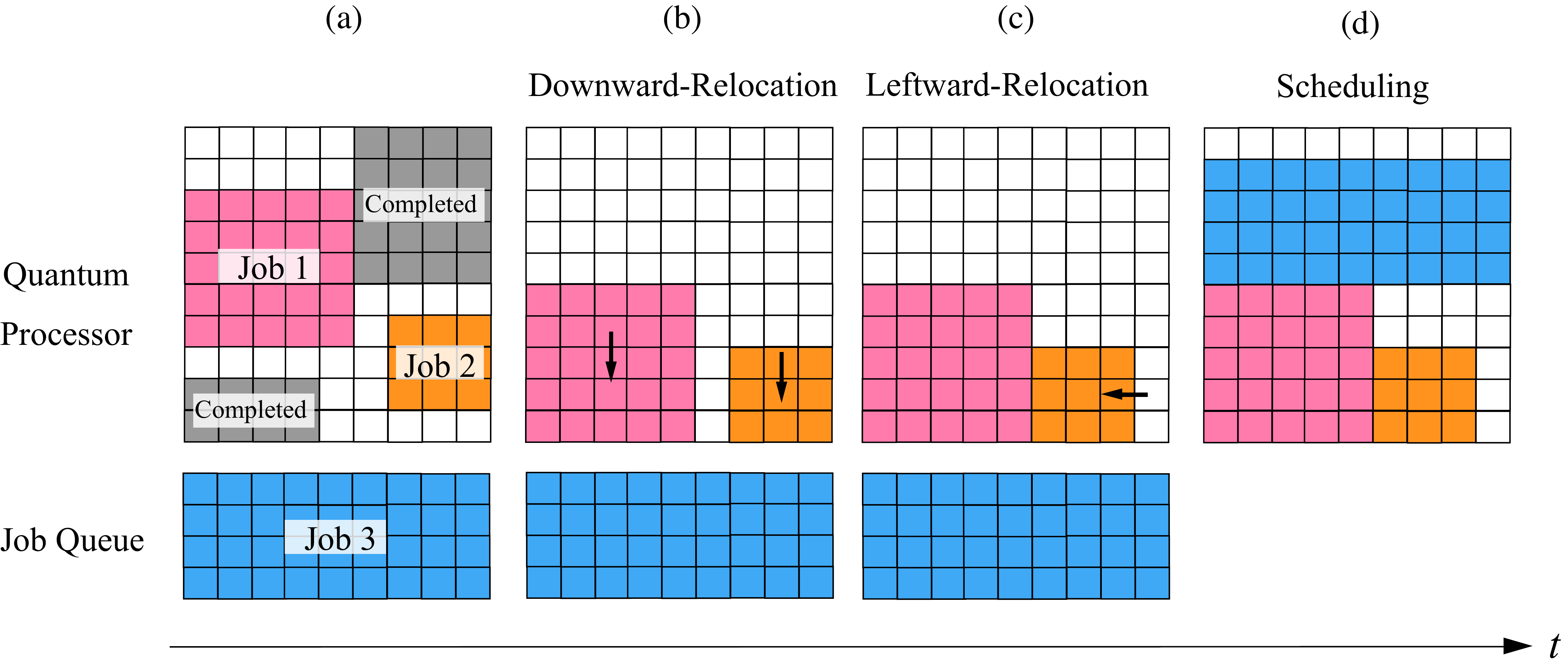}
    \caption{$XY$-plane views of defragmentation process. (a) Suppose that at some point in time, a processor is occupied by jobs, and there is a job queue. Two jobs have occupied some space but have now been completed. (b) First, defragmentation brings the running jobs (Job 1,2) to the top as much as possible. (c) Next, it pulls the jobs to the left as much as possible. (d) By utilizing the available space, a new job (Job 3) can be placed.}
    \label{fig:defrag}
\end{figure*}

In QMP on architectures with connectivity constraints, multiple jobs sharing the quantum chip can sometimes lead to fragmentation of the available chip space. For example, in \Cref{fig:defrag} (a), two of the four running jobs (indicated in gray) have completed execution. Although we want to allocate a new job 3 (indicated in blue), there is no sufficiently large contiguous space available, requiring job 3 to wait. However, as illustrated in \Cref{fig:defrag} (b) and (c), if we \emph{relocate} the positions of jobs 1 and 2, we can create enough space to accommodate job 3, allowing it to be scheduled without delay. 

To improve the utilization of quantum processors through relocation like \Cref{fig:defrag}, we incorporate the notion of \emph{defragmentation}~\cite{IBMdefrag} into our scheduler. The defragmentation algorithm is illustrated in \cref{alg:defrag}.  

\begin{figure}[!t]
  \begin{algorithm}[H]
    \caption{Defragmentation Algorithm}
    \label{alg:defrag}
    \begin{algorithmic}[1]
      \Require{$\mathcal{P}_{rsv}$ is the set of reserved jobs}
      \Require{$t$ is the timing to perform defragmentation}
      \Require{$I$ is the defragmentation interval}
      \Function{DefragAt}{$t$}
      \State $\mathcal{P}_{\text{below}} \gets \emptyset, \ \mathcal{P}_{\text{above}} \gets \emptyset$,
      \For{$P_i \in \mathcal{P}_{\text{rsv}}$}
        \State $(P_{\text{below},i}, P_{\text{above},i}) \gets$ Cut $P_i$ at $z = t$.
        \State Append $P_{\text{below},i}$ to $\mathcal{P}_{\text{below}}$
        \State Append $P_{\text{above},i}$ to $\mathcal{P}_{\text{above}}$
      \EndFor
      \State /* Shift $y$ positions */
      \State Sort $\mathcal{P}_{\text{above}}$ by $P.y_1$ for $P \in \mathcal{P}_{\text{above}}$
      \State $\mathcal{P}_{\text{tmp}} \gets \emptyset$
      \For{$P \in \mathcal{P}_{\text{above}}$}
        \State $y_{\text{max}} \gets 0$
        \For{$Q \in \mathcal{P}_{\text{tmp}}$}
          \State $\mathrm{flag}_x = \lnot (P.x_2 \leq Q.x_1 \lor Q.x_2 \leq P.x_1)$
          \State $\mathrm{flag}_z = \lnot (P.z_2 \leq Q.z_1 \lor Q.z_2 \leq P.z_1)$
          \If{$\mathrm{flag}_x \land \mathrm{flag}_z$}
            \State $y_{\text{max}} \gets \Call{max}{y_{\text{max}}, Q.y_2}$
          \EndIf
        \EndFor
        \State $P.y \gets y_{\text{max}}$
        \State Append $P$ to $\mathcal{P}_{\text{tmp}}$
      \EndFor
      \State /* Shift $x$ positions */
      \State Sort $\mathcal{P}_{\text{tmp}}$ by $P.x_1$ for $P \in \mathcal{P}_{\text{tmp}}$
      \State $\mathcal{P} \gets \emptyset$
      \State $\dots$
      \Comment{Similar procedure to shifting $y$}
      \State $S.\mathcal{P}_{\text{running}} \gets \mathcal{P}_{\text{below}} \cup \mathcal{P}$
      \EndFunction
      \State 
      \Function{Defrag}{$I$}
        \State $Z \gets \{P.z_2 \mid P \in \mathcal{P}_{\text{rsv}} \land P.z_2 > z_{\text{last}} \}$ \Comment{sorted in increasing order}
        \While{$|Z| > \text{threshold}$}
          \State $z_1 \gets$ $Z$.pop()
          \State $z_2 \gets$ $Z$.top()
          \If{$z_2 - z_1 \geq I$}
            \State \Call{DefragAt}{$z_1$}
            \State $z_{\text{last}} \gets z_1$
          \EndIf
        \EndWhile
      \EndFunction
    \end{algorithmic}
  \end{algorithm}
\end{figure}

In the proposed algorithm, all programs are assumed to be represented as cuboids. The algorithm begins by selecting a time \( t \) at which the cuboids will be reorganized. Among the reserved cuboids, those intersecting with \( z = t \) are split, and the cuboids located above \( z = t \) are relocated according to the following procedure: (1) each cuboid is first shifted downward along the negative \( y \)-axis as far as possible, and (2) then shifted left along the negative \( x \)-axis.

The cost of job relocation per defragmentation in our approach is at most \( W + H \) steps. This is because positions can be collectively shifted by rows or columns, and in lattice surgery, logical qubit movement can be performed in constant code cycles regardless of the distance to move. Since the execution time of practical FTQC programs is expected to be much longer than the chip size, it is expected that the overhead of relocation will not pose a major issue.

A distinguishing feature of our defragmentation approach is that a single cuboid can be divided into multiple parts. This contrasts sharply with traditional 3D bin packing problems, such as those found in logistics, where dividing a cuboid (e.g., a package) is not permitted.

\textbf{Defragmentation timing.}
The selection of candidate \( z \)-coordinates at which to perform defragmentation is crucial. In principle, it is natural to perform defragmentation when some job completes execution. Therefore, in this study, all time points at which previously scheduled jobs finish are considered candidate positions for defragmentation.

To effectively select a defragmentation point from among the candidates, we introduce a defragmentation interval parameter \( I \). In general, defragmentation should be performed in regions with low spatial utilization. In other words, regions with more significant gaps between consecutive candidate points are more likely to correspond to low-utilization periods. Therefore, we set \( I \) as a threshold parameter and perform defragmentation only at positions where the gap between consecutive points is at least \( I \).

Additionally, the invocation of the function \textsc{Defrag} is triggered at the beginning of each scheduling cycle. This is because, compared to the overall cost of scheduling, the defragmentation process itself is relatively lightweight.

\begin{figure}[tb]
    \centering
    \includegraphics[width=0.8\linewidth]{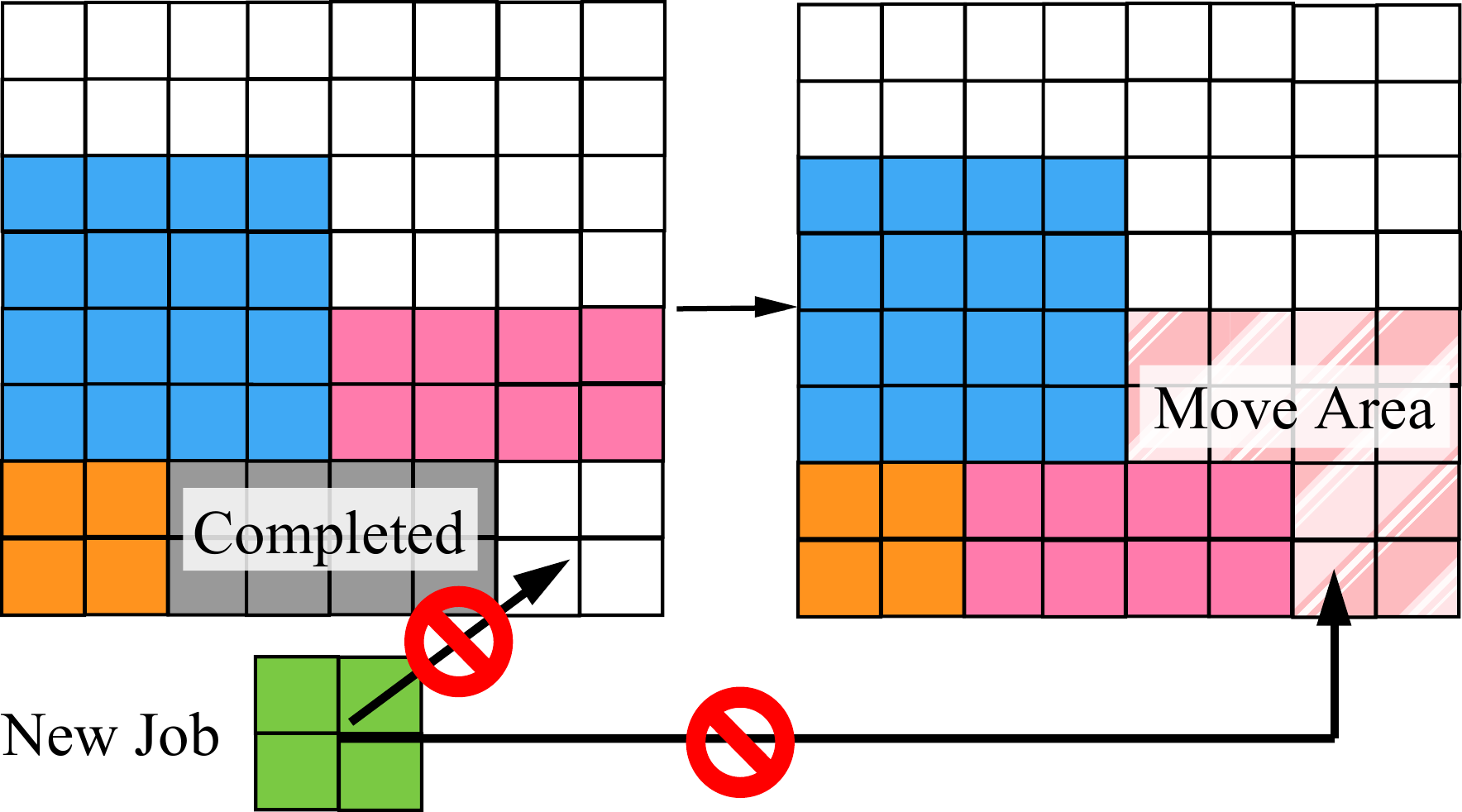}
    \caption{Placement constraints for new jobs introduced by defragmentation. Jobs cannot be placed in a manner that spans across the relocation regions used during defragmentation.}
    \label{fig:defrag-consistency}
\end{figure}

\textbf{Consistency of defragmentation.}
When performing defragmentation at a future time step, as in this study, it is essential to maintain \emph{consistency} with subsequent scheduling decisions. A defragmentation result is considered consistent if, at the designated time step, the planned relocations can be executed as intended. For instance, as illustrated in \Cref{fig:defrag-consistency}, no job can be assigned to overlap with the region through which another job is scheduled to move. Our defragmentation algorithm guarantees consistency by invoking \textsc{DefragAt} in chronological order and explicitly managing the relocation regions required for each move.

It should be noted that it is possible to consider defragmentation strategies that allow arbitrary relocation of the regions occupied by jobs. Even in scenarios with small ancillary space, such relocation can be achieved by permitting swap gates between jobs. While this approach has the potential to enhance the effectiveness of defragmentation, it may significantly increase both the cost of searching for an optimal placement and the overhead associated with relocation.
\section{Evaluation}\label{sec:evaluation}

To evaluate the performance of the proposed scheduler and the polycube-based scheduler~\cite{weko_226758_1}, we conducted a simulation-based performance analysis\footnote{The source code of our schedulers and experiments is available at \url{https://github.com/team-QMP/FTQMP-Scheduler}.}. In this evaluation, we aim to address the following research questions.

\begin{itemize}
\item \textbf{RQ1 (Throughput):} To what extent does the scheduler improve throughput? 
\item \textbf{RQ2 (Defragmentation):} How effective is defragmentation in improving scheduling performance?
\item \textbf{RQ3 (Responsiveness):} Does the scheduler operate with sufficient real-time responsiveness?
\end{itemize}

\subsection{Experiment Setup}

\Cref{tbl:parameters} summarizes the experimental parameters.
The latency of lattice surgery is proportional to code distance and code cycle.
We assumed the code cycle as 1 \si{\micro\second} with reference to typical values for superconducting QCs~\cite{battistel2023real}.
The minimum required code distance for quantum advantage has been estimated as $d=23$ in reasonable hardware assumptions~\cite{yoshioka2024hunting}. 
Thus, we used a code distance with a small margin, $d=31$, since we targeted a regime where sufficiently large FTQCs are provided as cloud services. 
In other words, program execution in our simulation advances by one voxel every \qty{31}{\micro\second}.
The defragmentation interval $I$ was fixed at $2 \times 10^4$ \si{\micro\second} for simplicity. 
However, it should be noted that in practical scenarios, the interval should either be configured based on the characteristics of the actual workflow or determined automatically by estimating appropriate defragmentation points.

\begin{table}[tb]
    \centering
    \caption{Parameters for Experiments}
    \label{tbl:parameters}
    \begin{tabular}{cc}
    \toprule
    Parameter & Value \\
    \midrule
    Code cycle & \qty{1}{\micro\second} \\
    Code distance $d$ & 31 \\
    Chip size $W \times H$ & $20 \times 20$ \\
    Batch size $B$ & 5 \\
    Defrag interval $I$ & $2 \times 10^4 \times d$ \si{\micro\second} \\
    \bottomrule
    \end{tabular}
\end{table}

To our knowledge, no benchmark instance has been proposed for FTQC-based QMP. 
Thus, we conducted experiments using randomly generated programs whose post-preprocessing cuboid sizes are summarized in \Cref{tbl:dataset-random}. 
Here, \([a, b]\) denotes that a value is chosen uniformly at random from the interval between \(a\) and \(b\), inclusive. 
For example, in Type {\gfour}, the values of \( w \) and \( h \) are chosen uniformly at random such that \( 10 \leq w, h \leq 20 \) and \( 101 \leq w \times h \leq 200 \). 
For simplicity, this study does not consider (shared) magic state factories or repeat-until-success constructs. 
For a discussion on handling these cases, please refer to \cref{sec:discussion}.

We prepared several classes of request data as shown in \Cref{tbl:dataset-job-requests}. 
Each class consists of 300 job requests, and we generated 50 instances for each class. 
All job requests were assumed to arrive at time $0$. 
All experiments were conducted on a classical computer system with a 13th Gen Intel(R) Core(TM) i7-1360P processor ($12$ cores, $16$ threads, Max Turbo Frequency $5.0$ GHz), and 32GB of RAM. We used CPLEX~\cite{cplex2009v12} to solve the ILP problems. 
For the ILP schedulers, the maximum allowed search time per schedule was set to 2 seconds.

\begin{table}[tb]
  \centering
  \caption{Dataset Information}
  \label{tbl:dataset-random}
  \begin{tabular}{cccc}
  \toprule
  Type & $w, h$ & $l$ & Constraint\\
  \midrule
  \gone   & $[5,10]$ & $[1 \times 10^4, 2 \times 10^4]$   &    \\
  \gtwo   & $[5,10]$ & $[4 \times 10^4, 6 \times 10^4]$   &    \\
  \gthree & $[5,10]$ & $[8 \times 10^4, 1 \times 10^5]$   &    \\
  \gfour  & $[10, 20]$ & $[1 \times 10^4, 2 \times 10^4]$ & $101 \leq w \times h \leq 200$ \\
  \gfive  & $[10, 20]$ & $[4 \times 10^4, 6 \times 10^4]$ & $101 \leq w \times h \leq 200$ \\
  \gsix   & $[10, 20]$ & $[8 \times 10^4, 1 \times 10^5]$ & $101 \leq w \times h \leq 200$ \\
  \bottomrule
  \end{tabular}
\end{table}

\begin{table}[tb]
  \centering
  \caption{Job Request Data Information}
  \label{tbl:dataset-job-requests}
  \begin{tabular}{ccc}
  \toprule
  Class & \#Request & Ratio of each data type \\
  \midrule
  A   & \multirow{9}{*}{300} & \gone : 50\%, others : 10\% each \\
  B   &                      & \gtwo : 50\%, others : 10\% each \\
  C   &                      & \gthree : 50\%, others : 10\% each \\
  D   &                      & \gfour : 50\%, others : 10\% each \\
  E   &                      & \gfive : 50\%, others : 10\% each \\
  F   &                      & \gsix : 50\%, others : 10\% each \\
  G   &                      & All: 16.6\% each (uniform ratio) \\
  H   &                      & \gone, \gtwo, \gthree : 30\% each, \gfour,\gfive,\gsix : 3.33\% each \\
  I   &                      & \gone, \gtwo, \gthree : 3.33\% each, \gfour,\gfive,\gsix : 30\% each\\
  \bottomrule
  \end{tabular}
\end{table}

\subsection{Experimental Results}

\subsubsection{RQ1. Throughput}

\begin{figure*}[htbp]
    \centering
    \includegraphics[width=0.7\linewidth]{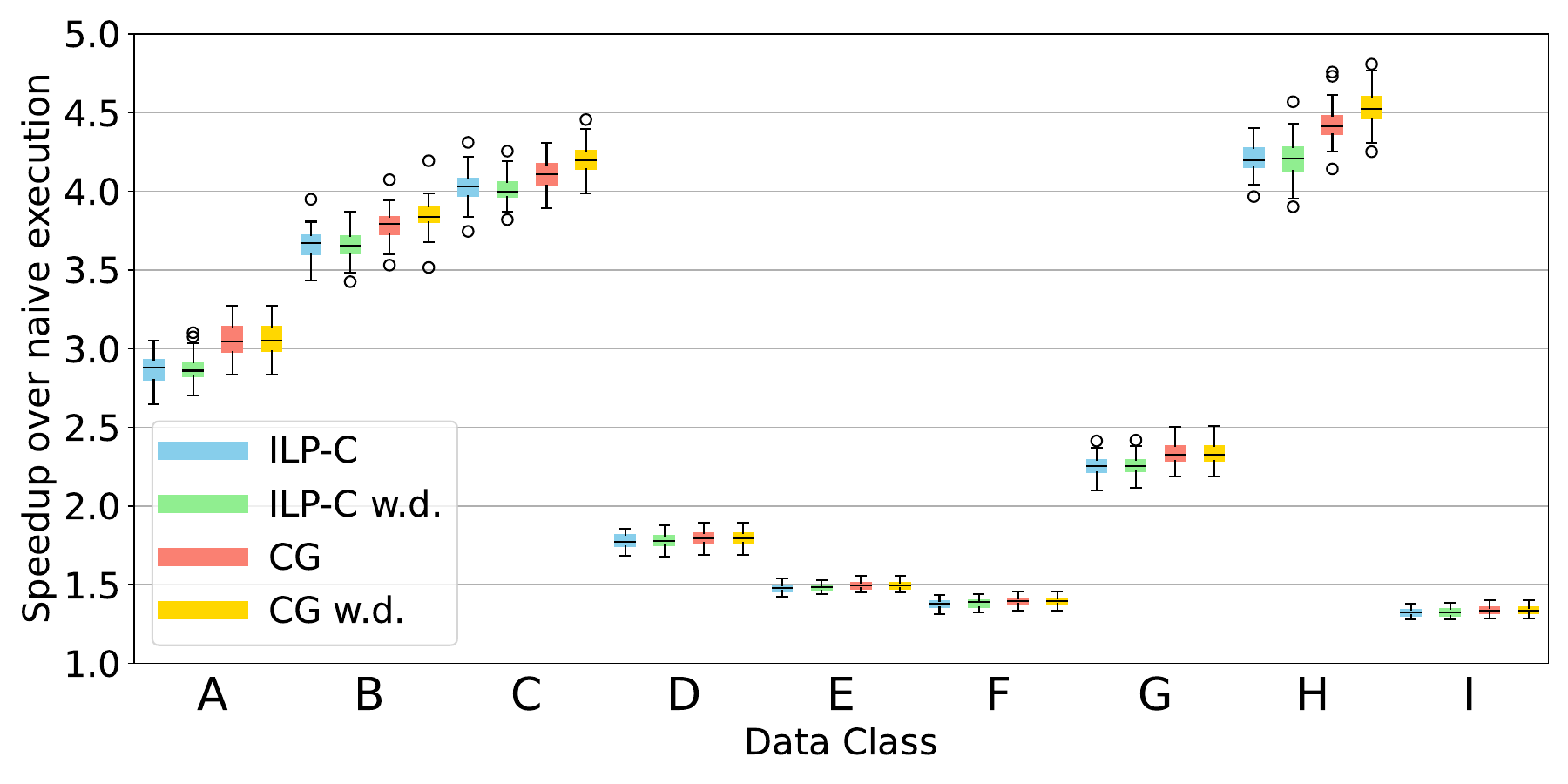}
    \caption{Speedup relative to naive total job execution time ($= \sum_i l_i$)} 
    \label{fig:results}
\end{figure*}

\begin{table}[tb]
    \centering
    \caption{Geometric mean of speedup across all classes}
    \label{tbl:geo-mean-all}
    \begin{tabular}{cccc}
    \toprule
    ILP-C & ILP-C w.d. & CG & CG w.d. \\
    \midrule
    2.32 & 2.32 & 2.38 & 2.40 \\
    \bottomrule
    \end{tabular}
\end{table}

\Cref{fig:results} shows the throughput results for each scheduler.
The vertical axis of the figure represents the speedup relative to $\sum_i l_i$, defined as the total job execution time when all jobs are executed sequentially.
%The experimental results are given in \Cref{fig:results} and \Cref{tbl:geo-mean-all}. 
In the figure, ``ILP-C'' refers to the ILP scheduler based on the cuboid representations, and ``CG'' denotes the corner greedy scheduler. 
For brevity, ``w.d.'' means ``with defragmentation''. 
\Cref{tbl:geo-mean-all} summarizes the geometric mean of the speedup achieved by each scheduler across all job classes.
%The ratio represents the extent of speedup achieved relative to the naive execution time of jobs $\sum_i l_i$.

The ILP scheduler based on the polycube representation~\cite{weko_226758_1} is excluded from \Cref{fig:results}, as it failed to produce feasible solutions within the time limit. 
This result highlights the severe scalability issues inherent in naive polycube-based scheduling approaches.

%%%%%%%%%%%%%% 2025/04/15 UENO changes
%For the dataset used in this study, the CG scheduler consistently outperformed the ILP-based scheduler across all job classes. 
%One contributing factor, as will be quantified in a later experiment, is the ILP scheduler's reduced responsiveness as the number of reserved jobs increases, leading to scheduling delays.

For the dataset used in this study, the CG scheduler consistently outperformed the ILP-based scheduler across all job classes. 
One contributing factor, as quantified in the later experiment, is the responsiveness of the ILP scheduler.
The latency of the ILP scheduler grows substantially as the number of reserved jobs increases, leading to scheduling delays and degraded performance.
%%%%%%%%%%%%%%%%%%%

Examining class-specific results, classes A, B, C, and H, where most jobs occupy relatively small chip areas, exhibited significant throughput improvements. 
Notably, for class H, the CG scheduler with defragmentation achieved a \( 4.53 \) times improvement in throughput. 
In contrast, the throughput did not improve significantly in classes D, E, F, and I. 
For example, in class I, the maximum observed improvement was limited to \( 1.34 \) times. 
These classes include a high proportion of jobs that occupy large areas of the chip, leaving a limited opportunity for improved packing density through scheduling.
Overall, the aggregated speedup across all classes, measured as a geometric mean, ranged from \( 2.3 \) to \( 2.4 \) times.
Among these, the corner greedy scheduler with defragmentation achieved the highest performance.

\subsubsection{RQ2. Defragmentation}

\begin{figure}[tb]
    \centering
    \includegraphics[width=1\linewidth]{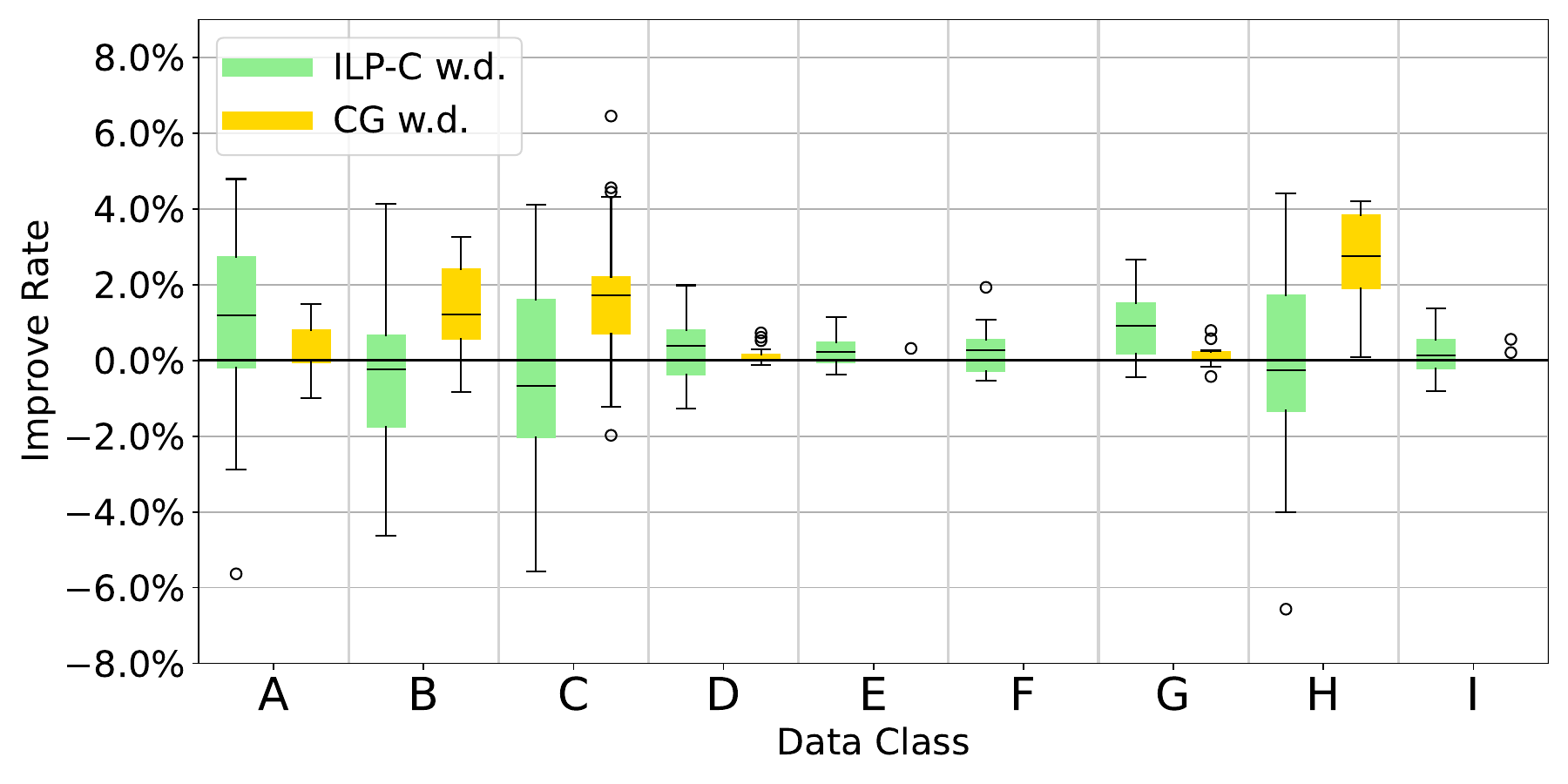}
    \caption{Improvement Rate (\%) through defragmentation against the scheduler without defragmentation}
    \label{fig:defrag-improve-results}
\end{figure}

Next, we investigated the throughput improvements achieved by defragmentation for each scheduler. 
The results are presented in \Cref{fig:defrag-improve-results}. 
Positive (negative) values in the figure indicate that defragmentation improved (degraded) throughput. 

Defragmentation is effective for the corner greedy scheduler, as represented by the yellow bars in \Cref{fig:defrag-improve-results}. 
Notably, in classes such as B, C, and H---where a substantial proportion of jobs occupy small chip areas for long durations and thus contribute to fragmentation---improvements of approximately 2\% were observed.

However, defragmentation does not always lead to performance gains; for example, the ILP solver results for classes B, C, and H are slightly degraded by introducing defragmentation. 
One possible reason is the need to reserve space for relocating job regions, as discussed in Sec.~\ref{sec:defrag}.
This requirement can reduce the availability of contiguous free space. 
This limitation could potentially be alleviated by incorporating defragmentation strategies that allow operations such as swapping job regions.

Another possible reason why defragmentation is not effective for the ILP-based scheduler lies in its formulation. 
Unlike the corner-greedy scheduler, the ILP scheduler does not explicitly aim to place jobs near the bottom-left corner of the chip. 
Since the defragmentation algorithm employed in this study assumes a bottom-left packing strategy, the ILP scheduler may produce more jobs that require relocation. 
Consequently, additional space must be reserved to accommodate these relocations, which in some cases may result in reduced overall efficiency.

%%%%%%%%%%%%%% 2025/04/15 UENO changes
%As demonstrated, the lightweight defragmentation algorithm proposed in this study can improve throughput. 
%However, performance may degrade due to constraints associated with relocation or mismatches between the defragmentation strategy and the scheduler's placement policy. 
%To achieve more consistent and higher improvement rates, it is essential to refine the relocation algorithm and pursue co-design of the scheduler and defragmentation mechanism.

In summary, as demonstrated by our experimental results, even our simple and lightweight defragmentation algorithm is effective in improving job scheduling to some extent, thereby presenting the potential of defragmentation for a high-performance FTQC job scheduler. 
To achieve more consistent and higher improvement rates, it is essential to refine the relocation algorithm and pursue co-design of the scheduler and defragmentation mechanism.
%%%%%%%%%%%%%%

\subsubsection{RQ3. Responsiveness}

Finally, we evaluated the responsiveness of the proposed schedulers, focusing on the time required to complete a single scheduling cycle under various batch sizes. 
For this experiment, we selected class G as a representative case, as it contains a balanced mix of all job types. 
To observe the scalability in response times, the time limit for the ILP solver was set to 30 seconds in this experiment.

\begin{table}[tb]
    \centering
    \caption{Batch Processing Time (\si{\micro\second}) per Scheduling}
    \label{tbl:responsiveness-result}
    \begin{tabular}{ccrrrr}
    \toprule
    Type & $B$ & Mean & Min & Max & StdDev \\
    \midrule
    ILP-C    & 5  & 244918 & 4962 & 14000017 & 831126 \\
    ILP-C    & 10 & 1646150 & 70221 & 21186320 & 1823412 \\
    ILP-C    & 15 & 5778808 & 553281 & 30102046 & 6098884 \\
    ILP-C    & 20 & 24384941 & 3230509 & 30086498 & 9427292 \\
    CG & 5  & 2652 & 13 & 15989 & 2652 \\
    CG & 10 & 5529 & 24 & 17444 & 4990 \\
    CG & 15 & 7948 & 50 & 25567 & 7190 \\
    CG & 20 & 10900 & 88 & 31623 & 9789 \\
    \bottomrule
    \end{tabular}
\end{table}

The experimental results are presented in \Cref{tbl:responsiveness-result}. 
The ILP scheduler exhibited a significant decline in responsiveness as the batch size increased, which can be attributed to the growing number of decision variables resulting from the increased number of cuboids considered simultaneously. 
By contrast, the greedy scheduler not only responded significantly faster than the ILP scheduler but also demonstrated a nearly linear scaling of response time with respect to batch size.

These results indicate that the ILP scheduler faces challenges in terms of real-time responsiveness compared to the greedy scheduler. 
Even for a batch size of \( B = 5 \), the average scheduling time reaches approximately \( 2.4 \times 10^5 \) \si{\micro\second}. 
This is nearly equivalent to the execution time of Type {\gone} and Type {\gfour} jobs, thereby diminishing the effective throughput improvement that scheduling is intended to achieve. 
This responsiveness issue is one of the reasons why the ILP scheduler does not outperform the greedy scheduler in overall performance. 
Furthermore, these results indicate that polycube-based schedulers, which have to account for more complex constraints, are not well-suited for use as online schedulers.

Nonetheless, it is important to note that the ILP scheduler may still be a viable option in scenarios where high responsiveness is not a primary requirement. 
Some estimates suggest that real-world workloads include applications with long execution times~\cite{yoshioka2024hunting}. 
Additionally, architectures such as those based on neutral atoms, which exhibit longer code cycle times compared to superconducting systems, may also be applicable. 
In such cases, the increased scheduling overhead may be justified by the potential to achieve more optimal placement configurations.
\section{Discussion}\label{sec:discussion}

In this section, we discuss the applicability of our approach to more practical scenarios that have not been considered in our schedulers, along with potential directions for extension. 

\subsection{Managing Shared Resources}
Some resources for FTQC, such as magic states and remote entangled states, are costly to generate. These resources may be produced in shared system assets and need to be allocated for each job.

How to \emph{fairly} distribute these valuable resources among jobs is a critical concern for future FTQC systems.

Furthermore, when supplying shared resources to a job, it is necessary to ensure that at least one accessible path exists between the resource factory and each job. One possible approach to guarantee this is to add \emph{frame qubits} before scheduling, as shown in Fig.~\ref{fig:surface_code}(c), which ensures access to the space surrounding the job.

\subsection{Addressing non-deterministic procedures}
The scheduling algorithm assumes programs start and end exactly at their scheduled times. However, quantum programs can have unpredictable execution times due to probabilistic protocols such as repeat-until-success. To handle this, it is possible to treat scheduling as approximate—focusing on the correct resource allocation order rather than exact timing. As long as each job starts after its resources are allocated, the scheduler remains effective despite timing variability.

\subsection{Towards Distributed QMP}
Distributed quantum computer (DQC)\cite{10.1063/5.0082975, 10.1145/3674151, PhysRevA.89.022317, 10.1145/1324177.1324179} is a promising candidate to realize scalable FTQC. Our work can be adapted to such systems with further consideration of the issues that emerge from the distributed environment. 

DQC requires the scheduler to minimize communication operations. Quantum circuit partitioning \cite{Omid2020Partitioning, Mahboobeh2020Evolutionary, burt2024generalised} can be employed by a scheduler. Quantum multiplexing, which enables the scheduler to perform multiple communications simultaneously, may also play an important role in preventing communication from becoming a system bottleneck~\cite{10.1117/12.893272, lo2019quantum, nishio2025multiplexed} because the communication channel can be a shared resource.

\subsection{Multi-job compiling}
Some NISQ-based QMP methods employ \emph{multi-job compiling}, which is a method for compiling multiple quantum circuits as a single circuit~\cite{liu2021QuCloudNew, ohkura2022simultaneous, niu2023enabling}. This approach allows for more flexible placement optimization by having the compiler, rather than the scheduler, perform the process of spatio-temporal placement of tasks, but it can cause an increase in server-side tasks and a reduction in responsiveness to users. Multi-job compiling is also applicable to FTQC and has the potential for more flexible scheduling when used in cooperation with our proposed scheduling process.
\section{Conclusion and Future Work}\label{sec:conclusion}
In this work, we formulated the online scheduling problem for FTQC-based QMP. 
Although we assumed that the quantum processor employs lattice surgery on surface codes, the scheduling is based on 3D space-time job requests and can therefore be straightforwardly generalized to any quantum programs on 2D quantum processors using different QECCs or logical gate implementations.

We proposed two schedulers: an ILP-based scheduler and a scheduler referred to as corner greedy. 
Both approaches rely on a preprocessing step that transforms polycube representations into cuboids, enabling significantly higher responsiveness compared to schedulers based directly on polycube formulations. 
Furthermore, by introducing the concept of defragmentation, we experimentally demonstrated that average throughput can be improved in some datasets.

We believe that incorporating advanced techniques from the field of online 3D bin packing could further enhance the schedulers. 
In addition, we plan to develop defragmentation strategies based on advanced relocation methods that permit inter-job swaps, along with accurate estimation of their associated costs. 
Other important directions for future work include incorporating fair allocation of shared resources, such as magic states, into the evaluation framework and extending the system to support DQC.

%We proposed a sub-optimal scheduler that we call \emph{Corner-Greedy} and evaluated the online scheduler with its throughput and responsiveness. Our scheduler outperformed the ILP-based scheduler in both throughput and responsiveness. 

%We also proposed \emph{preprocessing} for the scheduler which converts original polycube job requests into cuboids and $k$-split cuboids. They are bounding boxes for the polycube. These pre-processing steps improved the responsiveness of the scheduler and introduced the trade-off between responsiveness and the optimality of the schedule.

%In addition to the scheduling process, we proposed \emph{defragmentation} during the execution of jobs. It relocates running jobs as densely as possible to create space for adding new jobs. It was found that there are cases where this achieves \red{$n$-times higher throughput. To be updated with results.}

\section*{Acknowledgment}
We would like to thank Takeaki Uno and all members of the ``Software, Algorithms, and (Circuit) Optimization for Efficient Realization of Novel Device Architectures'' group for their invaluable advice regarding the formulation of the problem. SN would like to thank Kae Nemoto and Takahiko Satoh for useful discussions.
\bibliographystyle{unsrt}
\bibliography{references}

\end{document}